\documentclass{elsart}

\newcommand{\ignore}[1]{}

\usepackage{natbib}

\usepackage{graphicx}

\usepackage{amssymb}

\usepackage{lineno}

\linenumbers
\begin{document}

\begin{frontmatter}


\title{Simulations of the interaction \\
of cold gas with radio jets}
\author{Martin Krause}
\ead{M.Krause@mrao.cam.ac.uk}
\address{Astrophysics Group, Cavendish Laboratory,\\ 
	JJ~Thomson Avenue, Cambridge CB3 0HE, \\
	United Kingdom}


\begin{abstract}
A new scenario for the interaction of a jet with a background medium with 
cold clouds is investigated by means of hydrodynamic turbulence simulations
with cooling. The idea is that the cold clouds are overtaken by a radio 
cocoon and stirred up by turbulence in this cocoon.
The 2D multiphase turbulence simulations contain all the
three gas phases and have a number of interesting properties.
The produced power spectrum is proportional to the inverse square of the 
wavevector. The Mach number - density relation may explain the observed 
velocities in emission line gas associated with radio galaxies.
The model also explains the increased optical visibility within the
radio structures, the correlation between emission line and radio luminosity,
and the evolution of the alignment effect with source size. I also propose
this to be a useful model for the recently discovered neutral outflows in 
nearby radio sources.
\end{abstract}


\end{frontmatter}

\section{Introduction}\label{intro}
Extragalactic jets consist of relativistic, magnetised, radio-emitting plasma.
They are driven into the warm, usually X-ray-emitting also magnetised gas that 
surrounds galaxies. Colder gas components are observed in many systems as well. 
There is the {\em alignment effect}, i.e. optical emission lines and continuum
co-spatial with the area enclosed by the radio structures at redshifts beyond
$z \approx 0.5$ \citep{Longair2006}. Associated Lyman~$\alpha$ absorption
sometimes at the same redshift as molecular gas \citep{Krause2005b}
is detected at even higher redshifts.
Atomic and ionised hydrogen has also been detected in interaction with local 
radio sources \citep{Morgea2005a,Morgea2005b}.

Understanding the interaction of these components is not only an interesting 
piece of astrophysics in itself, but bears the potential to find out more
about the galactic nucleus that causes the jets, the environment of galaxies,
and the way jets influence the evolution of galaxies.

\section{Turbulence in jet cocoons}

The alignment effect describes the observation of cold ($<10^6$~K) gas associated
with radio structures. A common idea is that the cold gas is present in the
form of clouds, with low filling factor, which predate the jet phase.
Emission is thought to be produced either by shocks in these clouds
\citep{MKR02}, 
induced by the bow shock, or by photoionisation due to a hidden quasar
(there is also emission ascribed to newly formed stars).

A problem with this scenario is the sparsity of emission outside the 
radio structures \citep{Inskea02c}, where the clouds should still be 
photoionised by the quasar. The obvious alternative, cooling the cold
gas from the surrounding warm phase ($10^6$~K~$<T<10^8$~K) was so far 
excluded, since the cooling time of that gas, where observed, was much 
longer than the jet age, and also for dynamical reasons.

We have found a way to circumvent these difficulties. The new idea is that 
just a few preexisting cold clouds are overtaken by the jet. When they enter the
radio cocoon, together with surrounding warm gas, they are stirred up by the
violently turbulent plasma in the radio cocoon and form a turbulent system.
Now, turbulence transports energy between the phases, and the cold clouds
radiate it away. Hence, the whole system looses energy, as does the warm phase.
The turbulently enhanced cooling rate leads to cold mass condensation.
The increased cold mass is in turn responsible for an increased visibility
of optically emitting clouds, even when considering photoionisation
as ionisation mechanism.

\subsection{Simulations}

\begin{figure}
\centering
\includegraphics[width=0.45\textwidth]{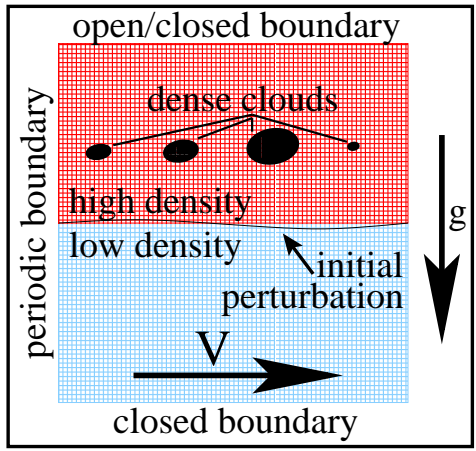}
\caption{Sketch of the setup for all simulations. The three phases soon mix
due to fluid instabilities, and form a turbulent region powered by the initial
gravitational and kinetic energy.}
\label{sketch}
\end{figure}

\begin{figure}
\centering
\includegraphics[width=\textwidth]{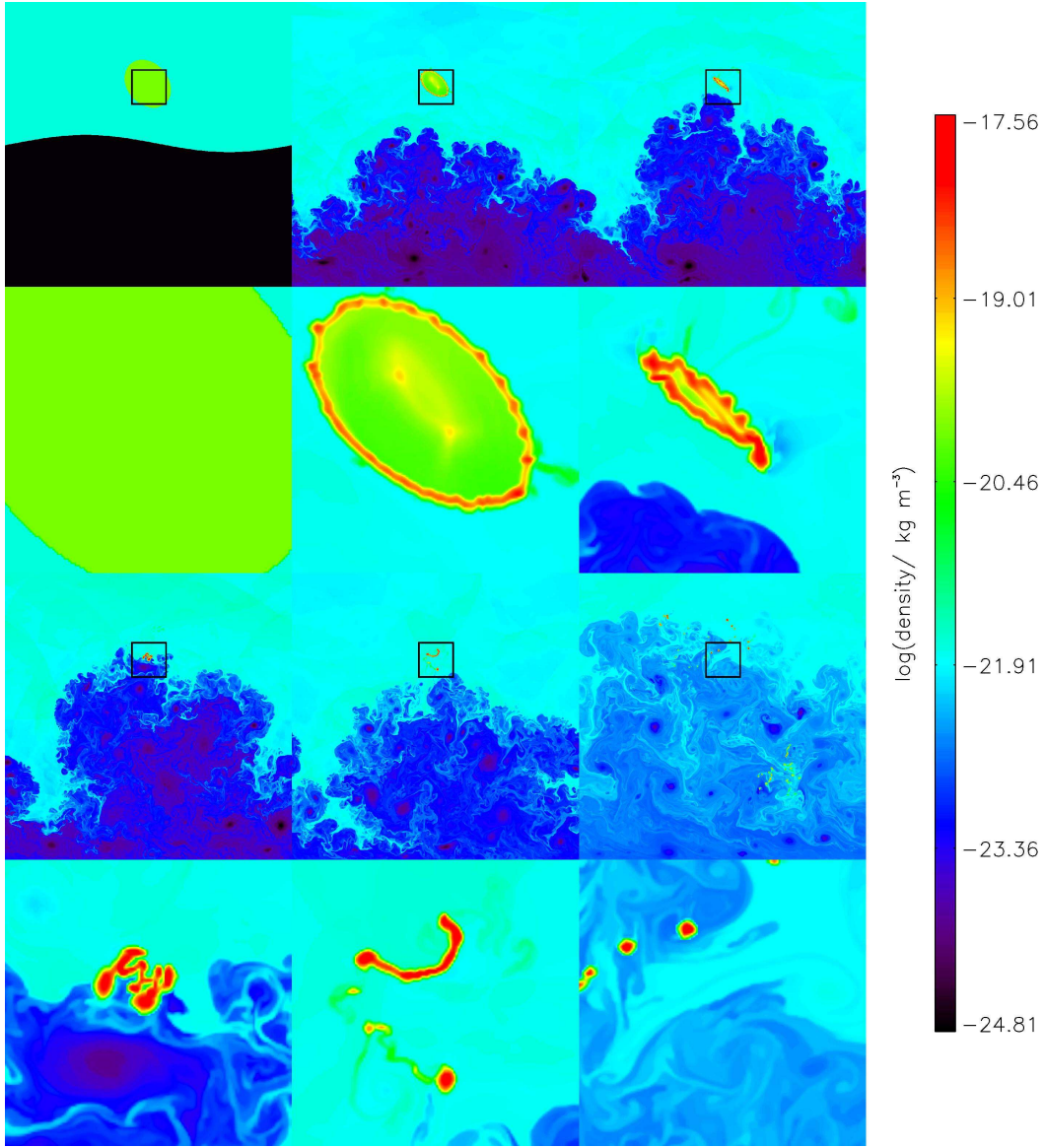}
\caption{Evolution of the logarithmic density for the Flash simulation.
	Time increases (0, 0.4, 0.5, 0.6, 0.8, 2.6 Myr) from top left
	to bottom right. Two plots are show at each time, the lower ones
	showing a magnification of the region marked with a box in the upper
	ones. The box marks always the same region.} 
\label{denevol}
\end{figure}

\begin{figure}
\centering
\includegraphics[width=.48\textwidth]{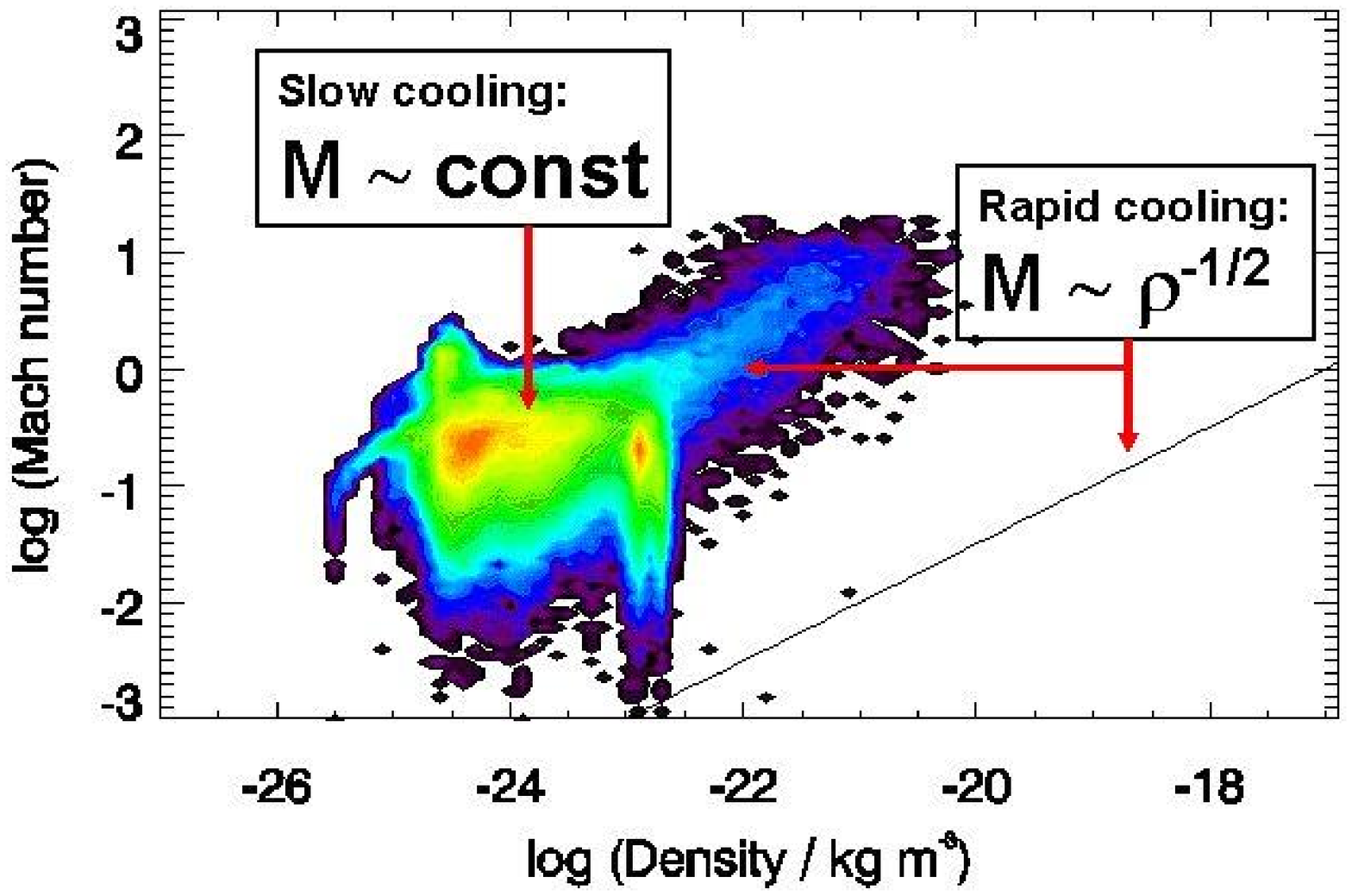}
\includegraphics[width=.48\textwidth]{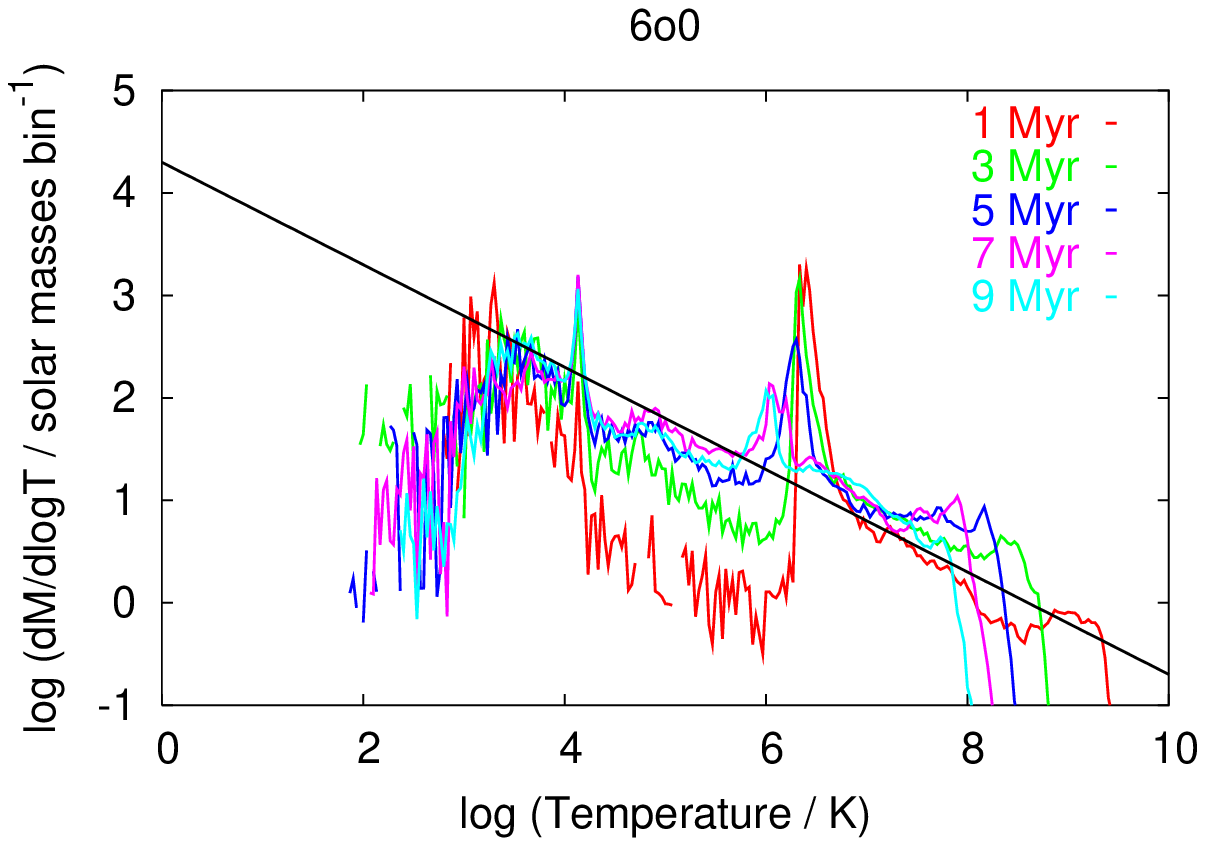}
\caption{Left: Typical Mach number - density histogram here for a Nirvana 
	simulation. Gas that is dense enough to cool on the simulation time
	has constant velocity ($M\propto \sqrt{\rho}$), non-cooling gas has 
	$\rho v^2=$~const ($M=$~const).
	Right: typical gas mass distribution versus temperature.
	The general $1/\sqrt{\rho}$ behaviour is probably due to mixing,
	The high temperature peaks reflect the initial condition, where the
	right cutoff moves quickly to lower temperatures again due to mixing.
	The peak around 14,000~K is produced by a dynamical equilibrium 
	between shock and compressional heating and cooling. It is located 
	at a sharp rise in the cooling function.}
\label{mdh}
\end{figure}

\begin{figure}
\centering
\includegraphics[width=.48\textwidth]{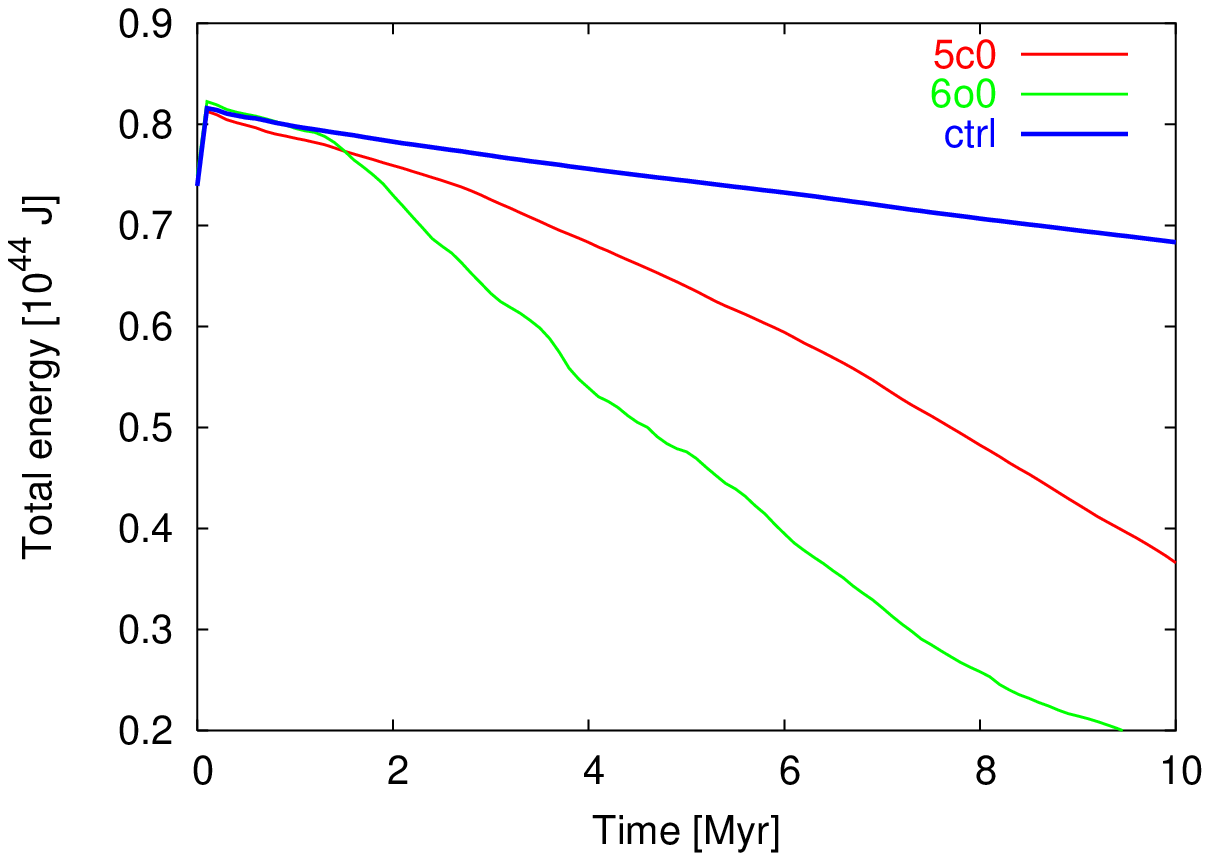}
\includegraphics[width=.48\textwidth]{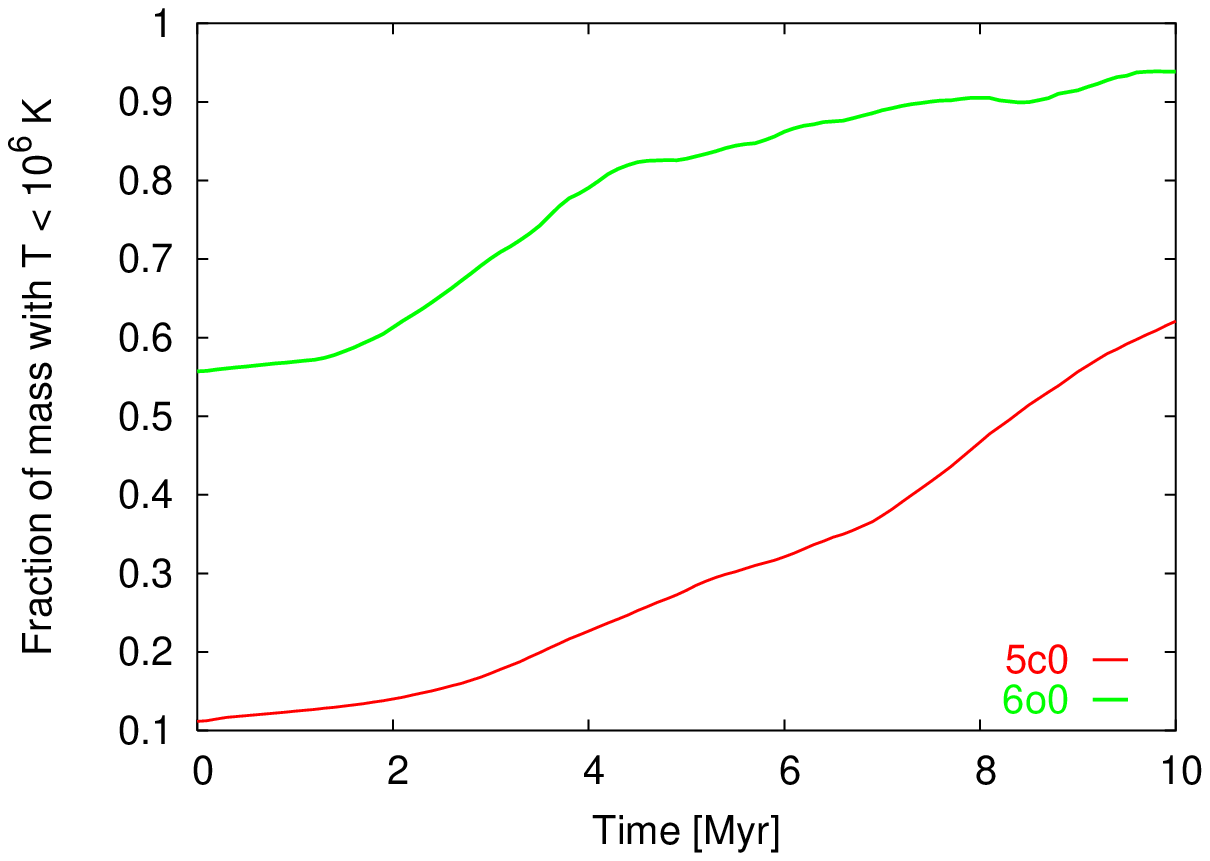}
\caption{Left: Energy loss in two Nirvana simulations with clouds and a control 
	run. The control run as well as the 5c0 simulation are closed box 
	simulations. Clearly, the cold clouds produce an enhanced system
	cooling rate. The clouds in the 6o0 simulation are an order of magnitude 
	denser, but it has an open upper boundary.
	Right: Cold gas mass fraction. The fraction of cold mass increases
	more rapidly with increasing cold mass content. This is not only true
	between the simulations but also for the time evolution of an individual
	one, which is an exponential growth till saturation at $>80\%$ of the
	total mass. The control simulation produces no cold mass at all for 
	all of the simulation time.} 
\label{loss}
\end{figure}

An idealisation of the proposed scenario is the Kelvin-Helmholtz instability 
with cool clouds in the denser phase. Since radio jet cocoons are usually less
dense than the surroundings, this may represent the situation at the contact 
surface between radio cocoon and shocked ambient gas with the expanding cocoon
just about to overtake some cool clouds. We have simulated this setup 
(see Fig.~\ref{sketch} for a sketch) with the two codes, Nirvana \citep{ZY97}
and Flash \citep{Fryxea00}. The physical model is hydrodynamics (continuity, 
momentum and energy equations) plus constant gravity and
optically thin cooling. We adopted the 
approach of \citet{BA03}, i.e. a cooling function which is non-zero at all 
temperatures due to molecules, atomic emission lines and bremsstrahlung in 
order of increasing temperature. Common sense suggests that the Flash
simulations should be more accurate due to the higher order of the 
interpolation scheme and the strict enforcement of energy and momentum
conservation. However, practical constraints (running time and the 
ability to cope with high density contrasts) made the Nirvana code the better
tool for this particular job. Also important: to have similar results with
different codes in particular with very different diffusivities gives more 
confidence in the conclusions.

We start with a 2D-Kelvin-Helmholtz setup with a density ratio of $10^4$ (Nirvana)
and $10^3$ (Flash). The less dense material has a tangential Mach number of 0.8.
Here I restrict myself to the discussion of cloud densities
of ten and a hundred times denser than the denser medium, and a control run
without clouds.

\subsection{Results}

The evolution of the density for the Flash run is shown in Fig~\ref{denevol}.
The cloud is shocked (the wobbling of the contact surface sends shocks into
the upper medium), compressed, torn apart, and forms little cloudlets that 
finally drift away from each other. The Nirvana simulations look quite similar
with the exception that they are, as expected, more diffuse and filamentary.

The turbulence we find is fairly isotropic with a power spectrum proportional 
to $k^{-2}$, where $k$ is the wavevector. This is what would be expected for 
uncorrelated shocks \citep{ES04}, and may be a special property of such 
multi-phase turbulence.

We always find a Mach number - density relation as shown in Fig.~\ref{mdh}
(left). For the low density, non-cooling part, the Mach number is constant. 
Because the pressure generally varies very little across the computational 
domain, this means that the kinetic energy density is independent of the 
density. The denser, cooling part has a Mach number proportional to the 
square root of the density. This means that the velocity is roughly constant 
in this regime. Hence, if the velocity in jet cocoons is of order light speed, 
and the density ratio to the cooling break in Fig.~\ref{mdh} of order $10^4$, as
suggested by other evidence \citep[e.g.][]{Krause2005a}, we derive velocities
of the stirred up cool gas of order 1000~km/s. This is indeed a figure that 
is often found for the emission line gas.

We identify a prominent peak in the mass distributions over temperature at about
14,000~K (Fig.~\ref{mdh}, right). This is not prescribed by the initial condition 
as the other peaks, but due to a dynamical equilibrium between the turbulent 
energy transfer to the cool clouds and the cooling. Unsurprisingly, it is
located at a sharp rise of the cooling function.

Energy loss and cold mass condensation increase drastically with increased 
cold mass load (Fig.\ref{loss}). The cold mass growth curve can be well fit
by an exponential. This is consistent with the naive expectation that the 
dropout rate is proportional to the amount of cold gas present.

\section{Discussion}

\subsection{Higher redshift radio galaxies}
As shown by Katherine Inskip at this meeting, emission line regions in connection
with higher redshift radio galaxies and the alignment effect are not always
within the region enclosed by the radio structures, which demonstrates the
complexity of the real situations. However, very often the gas is indeed 
within the radio structures. For such radio galaxies, 
the new mechanism proposed here may explain
a number observations that were so far less well connected by physical theories.
First, the observed emission line gas is steered up to high velocities, the 
order of magnitude of which can be explained by the Mach number - density
relation. Second, the mechanism should produce a correlation between emission
line and radio luminosity, since both is mainly powered by the turbulent energy
in the jet cocoon. This is a well known correlation in the literature \citep{MC93}.
The mechanism may explain the increased visibility of the cold, 
optically-emitting gas within the radio structures just by the fact that the 
cold mass suffers a drastic increase due to the turbulence enhanced cooling.
The observed evolution with source size (higher velocities and shock ionisation
in smaller sources) may be understood if one considers that turbulence decays
quickly. Turbulence in jet cocoons \citep[FR~II][]{FR74} is driven from the 
hotspots and decays
away from the hotspots. As long as the source is small, the hotspots are closest
to the galaxy where it can induce effectively multiphase turbulence. When 
the hotspots are further away, the turbulence has already decayed when reaching 
the central parts where most of the emission line clouds are located.
Therefore, the clouds are less stirred up, the velocity width is narrower,
and there is less energy available for shock ionisation. Finally, the scenario 
also supports jet induced star formation. This is because as the cool clouds
radiate their energy, they get more and more compressed and fragmented. 
The increased cloud density favours star formation.

The simulations have to be confirmed in 3D which I am working on 
currently. Another issue is that the size of the effect may depend on the 
temperature of the environmental gas. A change of the temperature of the gas 
surrounding the host galaxies might then explain why the alignment effect 
appears at $z>0.5$, only.

\subsection{Cold gas in nearby radio galaxies}
The jet-cold gas interactions reported by \citet{Morgea2005a} are probably 
produced by a different situation than the one simulated here.
However, since we end up with isotropic multi-phase turbulence, the model 
might be applicable in more general situations. The observations show ionised
and neutral hydrogen co-spatial with the radio cocoon of 3C-305. The radio 
morphology suggests that the jet is hitting a disc and knocking off some cold
gas. This seems to be accelerated to high outflow velocities, where both, 
neutral and ionised gas is observed. The simulations presented here
also predict a close connection between neutral and ionised gas. In total, 
when the turbulence is evolved, there is a similar amount of $T<10^4$~K gas
present as gas with $10^4$~K$<T<2\times 10^4$~K, which might be taken as proxy
for neutral and ionised hydrogen, respectively. The observations, however,
suggest about a hundred times more neutral than ionised gas. The source 
of the discrepancy might easily be with the simulations -- a more realistic
cooling function or 3D or maybe magnetic fields might make a difference.

A connection to absorbers in high redshift radio galaxies, as suggested 
elsewhere, seems unlikely to me. Those are quite narrow whereas the one
in 3C~305 is extremely broad. The jet in wind model I have proposed earlier
\citep{Krause2005b} seems not very appropriate here. The cold gas 
is probably not cooling behind a bow shock but entrained in a multiphase
flow away from the radio hot spots.

\section*{Acknowledgments}
MK acknowledges a fellowship from the Deutsche Forschungsgemeinschaft (KR 2857/1
-1)
and the hospitality of the Cavendish Laboratory, where this work has been 
carried out.
The software used in this work was in part developed by the
DOE-supported ASC / Alliance Center for Astrophysical Thermonuclear Flashes
at the University of Chicago.




\bibliographystyle{elsart-harv}
\bibliography{/home/krause/texinput/references}

\begin{thebibliography}{13}
\expandafter\ifx\csname natexlab\endcsname\relax\def\natexlab#1{#1}\fi
\expandafter\ifx\csname url\endcsname\relax
  \def\url#1{\texttt{#1}}\fi
\expandafter\ifx\csname urlprefix\endcsname\relax\def\urlprefix{URL }\fi

\bibitem[{{Basson} and {Alexander}(2003)}]{BA03}
{Basson}, J.~F., {Alexander}, P., Feb. 2003. {The long-term effect of radio
  sources on the intracluster medium}. \mnras 339, 353--359.

\bibitem[{{Elmegreen} and {Scalo}(2004)}]{ES04}
{Elmegreen}, B.~G., {Scalo}, J., Sep. 2004. {Interstellar Turbulence I:
  Observations and Processes}. Annu Rev Astron Astrophys 42, 211--273.

\bibitem[{{Fanaroff} and {Riley}(1974)}]{FR74}
{Fanaroff}, B.~L., {Riley}, J.~M., May 1974. {The morphology of extragalactic
  radio sources of high and low luminosity}. \mnras 167, 31P--36P.

\bibitem[{{Fryxell} et~al.(2000){Fryxell}, {Olson}, {Ricker}, {Timmes},
  {Zingale}, {Lamb}, {MacNeice}, {Rosner}, {Truran}, and {Tufo}}]{Fryxea00}
{Fryxell}, B., {Olson}, K., {Ricker}, P., {Timmes}, F.~X., {Zingale}, M.,
  {Lamb}, D.~Q., {MacNeice}, P., {Rosner}, R., {Truran}, J.~W., {Tufo}, H.,
  Nov. 2000. {FLASH: An Adaptive Mesh Hydrodynamics Code for Modeling
  Astrophysical Thermonuclear Flashes}. \apjs 131, 273--334.

\bibitem[{{Inskip} et~al.(2002){Inskip}, {Best}, {Rawlings}, {Longair},
  {Cotter}, {R{\" o}ttgering}, and {Eales}}]{Inskea02c}
{Inskip}, K.~J., {Best}, P.~N., {Rawlings}, S., {Longair}, M.~S., {Cotter}, G.,
  {R{\" o}ttgering}, H.~J.~A., {Eales}, S., Dec. 2002. {Deep spectroscopy of
  z\~{} 1 6C radio galaxies - I. The effects of radio power and size on the
  properties of the emission-line gas}. \mnras 337, 1381--1406.

\bibitem[{{Krause}(2005{\natexlab{a}})}]{Krause2005b}
{Krause}, M., Jun. 2005{\natexlab{a}}. {Galactic wind shells and high redshift
  radio galaxies. On the nature of associated absorbers}. \aap 436, 845--851.

\bibitem[{{Krause}(2005{\natexlab{b}})}]{Krause2005a}
{Krause}, M., Feb. 2005{\natexlab{b}}. {Very light jets II: Bipolar large scale
  simulations in King atmospheres}. \aap 431, 45--64.

\bibitem[{{Longair}(2006)}]{Longair2006}
{Longair}, M.~S., Aug. 2006. {The most luminous radio galaxies}. RevMexAA,
  Conf. Ser. 26, 101.

\bibitem[{{McCarthy}(1993)}]{MC93}
{McCarthy}, P.~J., 1993. {High redshift radio galaxies}. \aap Review 31,
  639--688.

\bibitem[{{Mellema} et~al.(2002){Mellema}, {Kurk}, and {R{\"
  o}ttgering}}]{MKR02}
{Mellema}, G., {Kurk}, J.~D., {R{\" o}ttgering}, H.~J.~A., Nov. 2002.
  {Evolution of clouds in radio galaxy cocoons}. \aap 395, L13--L16.

\bibitem[{{Morganti} et~al.(2005{\natexlab{a}}){Morganti}, {Oosterloo},
  {Tadhunter}, {van Moorsel}, and {Emonts}}]{Morgea2005a}
{Morganti}, R., {Oosterloo}, T.~A., {Tadhunter}, C.~N., {van Moorsel}, G.,
  {Emonts}, B., Aug. 2005{\natexlab{a}}. {The location of the broad H i
  absorption in 3C 305: clear evidence for a jet-accelerated neutral outflow}.
  \aap 439, 521--526.

\bibitem[{{Morganti} et~al.(2005{\natexlab{b}}){Morganti}, {Tadhunter}, and
  {Oosterloo}}]{Morgea2005b}
{Morganti}, R., {Tadhunter}, C.~N., {Oosterloo}, T.~A., Dec.
  2005{\natexlab{b}}. {Fast neutral outflows in powerful radio galaxies: a
  major source of feedback in massive galaxies}. \aap 444, L9--L13.

\bibitem[{{Ziegler} and {Yorke}(1997)}]{ZY97}
{Ziegler}, U., {Yorke}, H.~W., 1997. {A nested grid refinement technique for
  magnetohydrodynamical flows}. Computer Physics Communications 101, 54.

\end{thebibliography}





\end{document}